\documentclass[%
 reprint,
nofootinbib,
 amsmath,amssymb,
 aps,
onecolumn
]{revtex4-2}

\usepackage{float} 

\usepackage{mathtools}
\usepackage{graphicx}
\usepackage{dcolumn}
\usepackage{bm,color,stmaryrd}
\usepackage{hyperref}
\usepackage{upgreek}
\usepackage{siunitx}
\DeclareSIUnit\dyne{dyn}

\usepackage[dvipsnames]{xcolor}

\def\volfrac{\phi}
\def\nphi{\Phi}

\def\b{\boldsymbol}

\def\u{\b{u}}
\def\D{\b{\mathrm{D}}}
\def\I{\b{\mathrm{I}}}
\def\Q{\b{\mathrm{Q}}}

\def\bsigma{\boldsymbol{\sigma}}

\def\Er{\mbox{Er}_a}
\def\Act{\mathcal{A}}

\def\Teq{\tilde{\Theta}}

\def\p{\b{p}}

\def\x{\b{x}}
\def\m{\b{m}}
\def\n{\b{n}}
\def\h{\b{h}}
\def\H{\b{H}}
\def\u{\b{u}}

\def\Pe{\mathrm{Pe_s}}
\def\V{\mathcal{V}_0}
\def\W{\mathcal{W}}
\def\LL{\mathcal{L}}

\graphicspath{{figures/}}

\begin{document}

\preprint{APS/123-QED}

\title{Arrested development and traveling waves of active suspensions\\ in nematic liquid crystals}

\author{Jingyi Li$^1$}
\author{Laurel Ohm$^1$}%
\author{Saverio E. Spagnolie$^{1,2}$}
\affiliation{$^1$ Department of Mathematics, University of Wisconsin-Madison, 480 Lincoln Dr., Madison, WI 53706}%
\affiliation{$^2$ Department of Chemical \& Biological Engineering, University of Wisconsin-Madison
}%
\date{\today}

\begin{abstract}
Active particles in anisotropic, viscoelastic fluids experience competing stresses which guide their trajectories. An aligned suspension of particles can trigger a hydrodynamic bend instability, but the elasticity of the fluid can drive particle orientations back towards alignment. To study these competing effects, we examine a dilute suspension of active particles in an Ericksen-Leslie model nematic liquid crystal. An anchoring strength linking the active and passive media tunes the system between active suspension theory in Newtonian fluids in one limit, and active nematic theory in another. For extensile active stresses, beyond a critical active Ericksen number or particle concentration, the suspension first comes into alignment, then buckles via a classical bend instability. Rather than entering the fully developed roiling state observed in isotropic fluids, the development is arrested into a steady, flowing state by fluid elasticity. Arrested states of higher wavenumber appear at yet larger extensile activity, and a phase transition is identified at finite anchoring strength. If the active particles are motile, the particles can surf along the bent environment of their own creation. More exotic states are also observed, including a oscillatory `thrashing' mode.  Moment equations are derived, compared to kinetic theory simulations, and analyzed in asymptotic limits which admit exact expressions for the traveling wave speed and both particle orientation and director fields in the first arrested state.
\end{abstract}

\maketitle


\section{Introduction} \label{sec:Introduction}

Active particles in viscous fluids, from motile microorganisms to molecular-motor-driven biofilaments, generate stresses on the environment which can drive flow and affect particle orientational order. Biological fluids introduce additional complexity. Mucus, for instance, is anisotropic when sheared \cite{vhv93}, which can rectify and otherwise affect microorganism transport \cite{sp06,fd06,Chretien03,fdoa19}. Viscoelasticity and shear-dependent viscosity, generic in biological settings, also affect motility \cite{el15,lla21,Arratia22,su23}. Liquid crystals (LCs) \cite{dp93,Stewart19} have been used to probe the role of complex fluid stresses on bacterial transport in a more controlled setting \cite{zsla14,mttwa14,Lavrentovich16}. At high bacterial concentrations, collective behavior is observed \cite{Aranson22} which is distinguishable from such behaviors in isotropic viscous fluids.

In a Newtonian fluid, orientational alignment can emerge due to hydrodynamic interactions alone \cite{ss15}, but such uniformly aligned states are susceptible to a bend instability \cite{sr02,ss07,sk09}, leading to chaotic dynamics \cite{ss08,mjrlprs13,cvabdd20,os22}. Sheets of aligned active matter in Newtonian fluids are similarly unstable \cite{mess19}. When swimming through a nematic LC, however, microorganisms can be confined by elastic stresses to swim along the director axis \citep{lp09,zsla14,mttwa14,rmasw15,ptgwl16,gsla17,gcdll22,pgapf24}, depending on the anchoring conditions \cite{cpzba20}, which can affect the nature of the bend instability and the nonlinear dynamics so induced. 

Particle motility and/or the concentration of active particles in an otherwise passive medium introduce additional collective behaviors. Driven actin filaments \cite{swsfb10}, Quincke rollers \cite{bcddb13}, and swimming spermatozoa \cite{cppdcyd16} can all exhibit wave-like dynamics, even in isotropic fluids. Density waves in active media have been found in models ranging from the Viscek and Toner-Tu--type continuum models \cite{gc04,ndv07,cggr08,bm08,vz12,mjrlprs13,Chate20}, to active polar gels \cite{gm12} and models incorporating hydrodynamic particle interactions \cite{ls14,tk16}. Solitary waves have also been observed in these models and in experiments \cite{gregoire08, swsfb10}, where particle bands form at large particle concentration and system size.

Experiments using a dilute suspension of microorganisms swimming through LCs have shown a change from individual rectified motion to the formation of a wave-like jet of swimming particles \cite{tktgypwyad20}. The wave dynamics in this experiment suggest a balance between active and elastic stresses. The behavior of a LC hosting a suspension of active particles (a `living liquid crystal' \cite{zsla14}) thus depends on the particle concentration.

In related active nematics, the constituents provide both nematic order and active stresses \cite{drbhd15,gis17,dijys18}. Active nematics support steady streaming states and regions of hysteresis \cite{vjp05,mocy07,ey09,gmch11,gmch12,gis16,svjs23,pcra24}, traveling waves \cite{rj16}, and interfacial wave propagation \cite{gckym24}, which depend on the nature of stress generation (extensile or contractile), substrate interactions \cite{cym23,avpisd24} and curvature \cite{klsdgbmdb14,gtv17,nt20,blrp22,zsc25}, and flow aligning/tumbling properties. These steady streaming states, in which fluid elasticity arrests further instability development, were studied recently by Lavi et al. \cite{lajc24}. High wavenumber labyrinthine patterns emerged at large active stresses. Pseudo-defect formation was also explored, and defects were found to destabilize arrested states into active turbulent states. A natural question arises: in what settings might such states emerge in living liquid crystals? Along these lines, a first order phase transition of a nematic LC due to active (kinesin/tubulin) interfacial stresses has recently been observed \cite{bmnnsm24}.

In this paper, we present a model for a dilute suspension of active particles in a nematic liquid crystal. The model ranges from an active suspension theory in either Newtonian \cite{ss08} or transversely isotropic environments \cite{hcsgcd18}, to an active nematic theory, controlled by an anchoring strength which promotes alignment between the active and passive media. An active Ericksen number, comparing active stresses to elastic fluid stresses, is used to explore a wide range of related systems. We observe bifurcations in the dynamics as either the particle concentration or the active Ericksen number is increased for extensile-stress-generating (`pusher') particles, first from uniform alignment and no flow, to the development of an arrested state with steady streaming, and then to arrested states of higher wavenumber. If the particles are motile, arrested states can translate at a finite wave speed, or even express a periodic `thrashing' mode. Analytical estimates are provided throughout.

\section{Mathematical modeling} 

\subsection{Active particle distribution and moments}
The active particles are modeled as prolate ellipsoids with major and minor axis lengths $2a$ and $2b$, volume $v=4\pi ab^2/3$, and surface area $S$, residing in a cubic domain with volume $L^3$. The number of particles, $N$, is assumed constant, and $\volfrac = Nv/L^3$ is the particle volume fraction. The number density of particles is written as $\psi^*(a\x, \p, T t)=NL^{-3}\psi(\x, \p,  t)$, where $a\x$ is the spatial position, $\p$ is the particle orientation, and $Tt$ is time ($T$ is defined below). Here, $\x\in[0,\LL]^3$ with $\LL=L/a$, and $\p\in \mathbb{S}^2$ with $\mathbb{S}^2$ the unit sphere. The total number of particles in the system is given by
\begin{equation}
    N = \int_{\mathcal{D}}\int_{\mathbb{S}^{2}} \psi^*(a\x, \p, T t) \,d\Omega  (a^3dV),
\end{equation}
where $a^3dV$ and $d\Omega$ are the infinitessimal volume element and solid angle, respectively, and $\mathcal{D}$ is the dimensionless fluid domain with $|\mathcal{D}|=\LL^3$. In dimensionless terms, then,
\begin{gather}
1 =\LL^{-3}\int_{\mathcal{D}}\int_{\mathbb{S}^{2}} \psi(\x, \p, t) \,d\Omega  \,dV.
\end{gather} Note that a uniformly concentrated, isotropic suspension has $\psi^*=N/(4\pi L^3)$, or $\psi=1/4\pi$. The particle concentration is written as $c^*(a\x,Tt)=N L^{-3}c(\x,t)$, where
\begin{gather}
    c(\x,t) = \int_{\mathbb{S}^{2}} \psi(\x,\p, t) \,d\Omega,
\end{gather}
so that a uniform concentration corresponds to $c(\x,t) = 1$. The first two orientational moments are denoted by
\begin{gather}
     \langle \p \rangle :=\int_{\mathbb{S}^{2}} \p \,\psi(\x,\p,t) \,d\Omega,\,\,\,\,
     \langle \p\p \rangle :=\int_{\mathbb{S}^{2}}  \p\p \,\psi(\x,\p,t) \,d\Omega,
\end{gather}
respectively, with $\p\p$ a dyadic product, and are normalized as
\begin{gather}\label{eqn:m_and_Q}
    \m(\x,t):= \frac{\langle \p \rangle}{c(\x,t)},\,\,\,\,
    \D(\x,t):= \frac{\langle \p\p \rangle}{c(\x,t)}.
\end{gather}
Here $\m$ is the polar order parameter for the active particles, and $\D$ relates to the traceless nematic order parameter, $\Q$, via $\Q=(\D-\I/3)$ \cite{dp93}.

\subsection{Ericksen-Leslie theory}

The active particles are assumed to be much larger than the LC molecules - a Type VI system in the language of Ref.~\cite{su23}. The LC dynamics are treated using the Ericksen-Leslie model (in the one-constant approximation) \citep{ll86,dp93}. Denoting the local LC orientation by $\n$, the bulk energy density is modeled as $(K/a^2)\mathcal{F}$, where (see \cite{suppmat})
\begin{gather}\label{Eq:F}
    \mathcal{F} = \frac{1}{2} \|\nabla  \n \|^2 + \frac{\W}{2}\int_{\mathbb{S}^2}\psi(\x, \p, t)\left(1-(\n \cdot \p)^2\right) \,d\Omega.
\end{gather}
Tangential anchoring conditions of strength $W$ are assumed on the LC/particle boundaries \cite{cpzba20}, resulting in the energetic penalty on misaligned particles above, with $\W=SW/(aK)$ the dimensionless anchoring strength \cite{bd70,lhsfrl04,tmml14} and $S\approx \pi^2 ab$ for slender particles with $b/a \ll 1$. Variations in the molecular position and orientation produces a stress $(K/a^2)\bsigma_r(\x,t)$, where
\begin{equation}\label{eqn:sigmar}
    \bsigma_r = -\nabla \n \cdot \nabla \n^T - \frac{\lambda}{2}(\h \n+\n \h)+\frac{1}{2}(\h \n-\n \h),
\end{equation}
with $\h = (\I -\n\n)\cdot \H$. Here, $\H=-\delta \mathcal{F}/\delta \n = \left[ \nabla^2 \n + \W \n\cdot (c \D) \right]$ is the molecular field, $\lambda$ is the tumbling parameter, and $\h\n$ and similar terms are dyadic products \cite{suppmat,ll86,dp93}.

Each particle imposes a force dipole of strength $\sigma $ on the surrounding fluid, resulting in a locally averaged active stress $\sigma  c\D$ \cite{hs11,ss15}. Defining now the timescale $T=\mu a^2/K$, the fluid velocity is denoted by $\u^*(a\x,Tt)=K(\mu a)^{-1}\u(\x,t)$, where $\mu$ is the solvent viscosity (anisotropic viscosity coefficients are neglected \cite{suppmat}). Momentum balance is given in dimensionless variables by
\begin{gather}
    -\nabla p+\nabla^2\u+\nabla \cdot\left(\bsigma_r+\Er \volfrac c \D\right)=\b{0},\label{eqn:Stokes}
\end{gather}
with $p$ the pressure, a Lagrange multiplier which enforces continuity, $\nabla \cdot \u=0$, and $(\nabla \cdot \bsigma)_j=\partial_i\sigma_{ij}$ (note the choice of convention, which is important since $\bsigma$ is not generally symmetric). Here we have defined the active Ericksen number, $\Er=\sigma  a^2/(vK)$, which scales with the active stress and inversely with LC elasticity~\cite{gbmsc14}. If $\Er<0$, the particles are extensile `pushers', and with $\Er>0$, they are contractile `pullers'. 

The director field dynamics are given by
\begin{equation}
    \frac{\mbox{D} \n}{\mbox{D} t} = (\I -\n \n) \cdot \left[ \n \cdot (\lambda \mathbf{E}+\mathbf{\Omega}) + \frac{1}{\gamma} \h \right] = (\I -\n \n) \cdot \left[ \n \cdot (\lambda \mathbf{E}+\mathbf{\Omega}) + \frac{1}{\gamma} \left(\nabla^2\n+\W c \n\cdot \D\right) \right], \label{eqn:LCdir}
\end{equation}
where $\gamma$ is the dimensionless rotational viscosity, $\mathbf{E} = (\nabla\u +\nabla \u^T)/2$, and $\mathbf{\Omega} = (\nabla\u -\nabla \u^T)/2$ \cite{suppmat,ll86,dp93}. 

\subsection{Active suspension dynamics}
Conservation of active particle number is expressed by a Smoluchowski equation for $\psi$,
\begin{equation}
    \psi_{t}+\nabla \cdot(\dot{\x} \psi)+\nabla_{\p} \cdot(\dot{\p} \psi)=0, \label{eqn:AMpsi}
\end{equation}
(see \cite{doi88}), where $\nabla_{\p} = (\I -\p\p)\cdot (\partial/\partial \p)$, and $\dot{\x}$ and $\dot{\p}$ are particle translational and rotational velocities, modeled as
\begin{gather}
    \dot{\x} =\u+\V \p-D\displaystyle\frac{\nabla \psi}{\psi}, \\
    \dot{\p} =(\I -\p \p) \cdot \left[ \p \cdot \nabla \u + \displaystyle\frac{\W}{\eta}(\n \cdot \p) \n\right]-d \displaystyle\frac{\nabla_{\p} \psi}{\psi}.\label{eqn:flux_p}
\end{gather}
With $V_0^*$, $D^*$, and $d^*$ the dimensional swimming speed, translational diffusivity, and rotational diffusivity, their dimensionless counterparts used above are $\V  = \mu a V_0^*/K$, $D=\mu D^*/K$, and $d=\mu a^2 d^*/K$. The moment acting on the LC by misaligned particles is balanced above by a corresponding torque $-S W  \n\cdot \p$ on each particle, affecting orientation via a near-alignment rotational drag with coefficient $\eta$. In a Newtonian fluid, $\eta\sim 16\pi(6\log(2a/b)-3)^{-1}$ as $b/a\to 0$. We also define $\Pe:=\V/D=a V_0^*/D^*$, a swimming Péclet number, and note that other terms can alternatively be written in terms of various Péclet and Ericksen numbers (e.g. $D=\Er/\mathrm{Pe}_a$, where $\mathrm{Pe}_a=\sigma a^2/(\mu v D^*)$ is an active Péclet number). 

The system is governed by the dimensionless groups $(\LL,\Er,\W,\V ,\gamma,D,d,\eta,\lambda)$. Characteristic values may be estimated using {\it B. subtilis} cells in Disodium cromoglycate (DSCG), a lyotropic chromonic LC \cite{mttwa14,zsla14,szla15,ztgsal17,tktgypwyad20}. Values associated with bacteria as the active particles are: $a \approx3\mu$m, $b\approx0.5\mu$m,  $S\approx15\mu$m$^2$, $v=4 \pi a b^2/3=3.1\mu$m$^3$, and $V_0^*\approx20\mu$m/s, from which we estimate $\eta \approx 4$, $f\approx 6\pi\mu a V_0^*\approx 10^{-7}$dyn and $\sigma=2af\approx 6\cdot 10^{-7}$dyn\,$\mu$m. LC material constants are approximately  $K=10^{-7}-10^{-6}$dyn, $W\approx 10^{-8}-10^{-7}$dyn/$\mu$m, and $\gamma^*=10^{-6}-10^{-4}$dyn s/cm$^2$ \cite{znnbsls14,ycllhbt21}. Then, with thermal energy $k_b T=4.11\cdot 10^{-21}\mbox{J}$, we have $D^*\approx  k_b T/(6\pi\mu a)=10^{-9}$cm$^2$/s and $d^*\approx  k_b T/(8\pi\mu a^3)=0.01$s$^{-1}$. Finally, assuming a domain with dimensions $\LL=50$, we then find the approximate dimensionless quantities $|\Er| \approx 1-20$, $\W\approx 0.5-5$, $\V \approx 0.06$, $\gamma\approx 10^{-4}$, $D, d\approx10^{-5}$, and we fix $\lambda=1$.

\begin{figure*}
\includegraphics[width=1\textwidth]{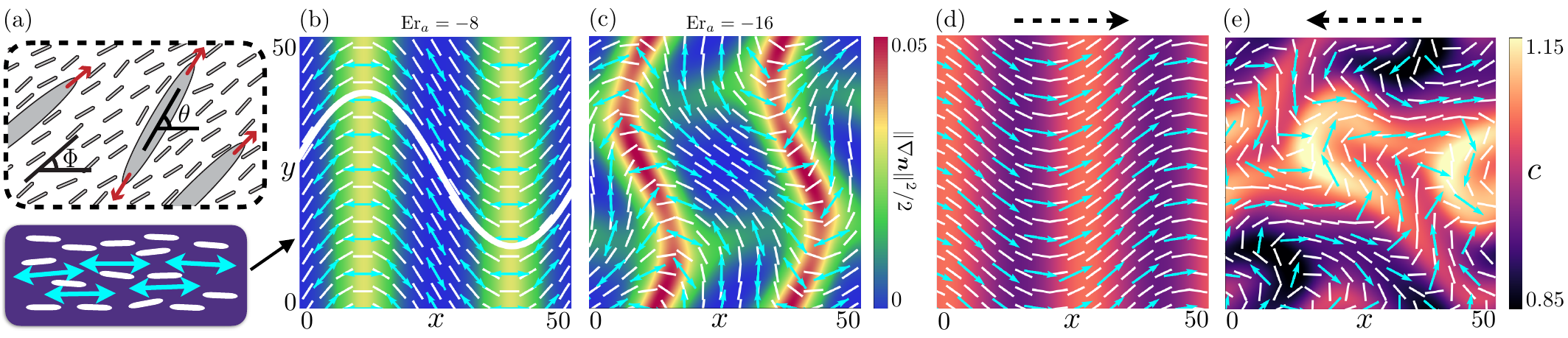}
\caption{(a) Schematic: individual elongated particles with orientations $\p = (\cos \theta, \sin \theta,0)$ are immersed in a nematic liquid crystal (LC) with director field $\n = (\cos \nphi, \sin \nphi,0)$. (b) The first arrested flowing state ($\Er = -8$, $\W=20$) from simulations using the full kinetic theory. White lines show the LC direction, cyan double arrows show the active particle orientations, and the background color shows the local LC elastic energy, $\|\nabla \n\|^2/2$. A director line (integral curve of the active particle field) is also shown. The fluid velocity is everywhere in the vertical direction, and largest in magnitude in the regions of large LC bending (i.e. active flow drives local bending and high elastic energy). See Movie 2. (c) The second arrested state  ($\Er = -16$) is fully two dimensional, following a secondary instability. See Movie 3. (d) Motile particles (here $\V =0.06$ and $\Er = -3$) relax into a rightward traveling concentration wave. Cyan arrows show the direction of particle swimming, and the background color shows the particle concentration. Particles evacuate from highly bent regions and cluster away from them. See Movie 4. (e) A leftward traveling wave of greater elastic deformation, with $(\Er, \volfrac, \V) =(-7, 0.1, 0.06)$, which moves in the opposite direction as the mean particle swimming direction (here on average directly to the right). See Movie 5.} 
\label{Figure1}
\end{figure*}

\section{Numerical simulations}  

We consider confinement to motion in two dimensions $(x,y)$ and invariance in the third dimension, writing $\p = (\cos\theta, \sin \theta,0)$ and $\n = (\cos\Phi, \sin \Phi,0)$ (Fig.~\ref{Figure1}a). Simulations of Eqs.~\eqref{eqn:Stokes}-\eqref{eqn:AMpsi} are performed using a pseudospectral method with $128^3$ Fourier modes and dealiasing; time-stepping is performed using an integrating factor method, along with a second-order accurate Adams-Bashforth scheme \cite{Fornberg98,mess19}. 

We begin by studying immotile particles $(\V=0)$ in the limit of small active Ericksen number, where elastic stresses dominate active stresses. For generic initial conditions in both particle concentration and orientation, the system rapidly stabilizes to a uniform LC orientation with active particle alignment in the LC direction. Diffusive spreading then leads to a constant concentration field and zero fluid velocity (see Movie 1).

Beyond a critical active Ericksen number or particle concentration, the classical bend instability of active particles overcomes the stabilizing LC elasticity. With a nearly uniform concentration of active particles in near-alignment with a uniform LC orientation field, a long-wave instability is triggered in both the particle and LC orientation fields. However, rather than continuing on to a fully developed roiling state like those seen in active suspensions in Newtonian fluids \cite{ss08,os22} and in active mixtures \cite{ybm20,sag20}, LC elasticity arrests further growth and the system relaxes to a steady flowing state (as in Refs.~\cite{vjp05,mocy07,gml08,zsla14,dbysjps18,skvgme19,lvrtbd23,lajc24}), evoking the classical Fr\'eedericksz transition in nematic LCs \cite{dp93,Stewart19}. 

Figure~\ref{Figure1}b (Movie 2) shows such a steady flowing state, with background color indicating the local LC elastic energy, $\|\nabla \n\|^2/2$, computed using $(\Er,\volfrac,\W,\gamma, \eta, \lambda, D, d) = (-8, 0.02, 20,0.1, 1, 1, 1, 0.1)$ and random initial data. The velocity field is everywhere vertical and fastest in the region of large LC bending. An integral curve of the director field (a director line) is included, for which an analytical expression will be derived in the following section.

For larger active Ericksen numbers, this arrested state succumbs to a secondary instability in the transverse direction, just as in the Newtonian setting \cite{ss15,fwz15,os22}. Here the arrested state takes on a fully two-dimensional configuration (Fig.~\ref{Figure1}c, Movie 3, where $\Er=-16$). Elastic stress in the LC becomes more focused into highly bent regions, and fluid flows with high velocity along these ridges (bending the LC most strongly there). The particle concentrations in both examples above equilibrate to uniformity, $c=1$.

For motile particles, arrested states again form but are found to translate at a speed which depends on the active Ericksen number. Figure~\ref{Figure1}d shows a traveling arrested state using the same parameters as those used in Fig.~\ref{Figure1}b except for $(\Er,\V )=(-3,0.06)$. The background color shows the attracting state of particle concentration, with waves moving to the right, in the direction of average particle motion (see Movie 4). Particles evacuate from regions of large bending, and cluster just ahead of the bent region, which moves along with them.

Surprisingly, the direction of the traveling wave can reverse relative to the individual particle swimming direction. Fig.~\ref{Figure1}e (and Movie 5) show the traveling arrested state with instead $(\Er, \volfrac)=(-7, 0.1)$. The concentration wave moves to the left, while the mean swimming direction is still directly to the right. Such a prograde-retrograde reversal has also been found in undulatory locomotion in nematic LCs \cite{ksp15}. At even higher swimming speeds and activity, a periodic `thrashing' mode appears. This mode is characterized by dramatic shifting of concentration and elastic energy back and forth between two symmetric states similar to Fig.~\ref{Figure1}e and its reflection across the $x$-axis (see Movie 6). 

\begin{figure}
\includegraphics[width=1\textwidth]{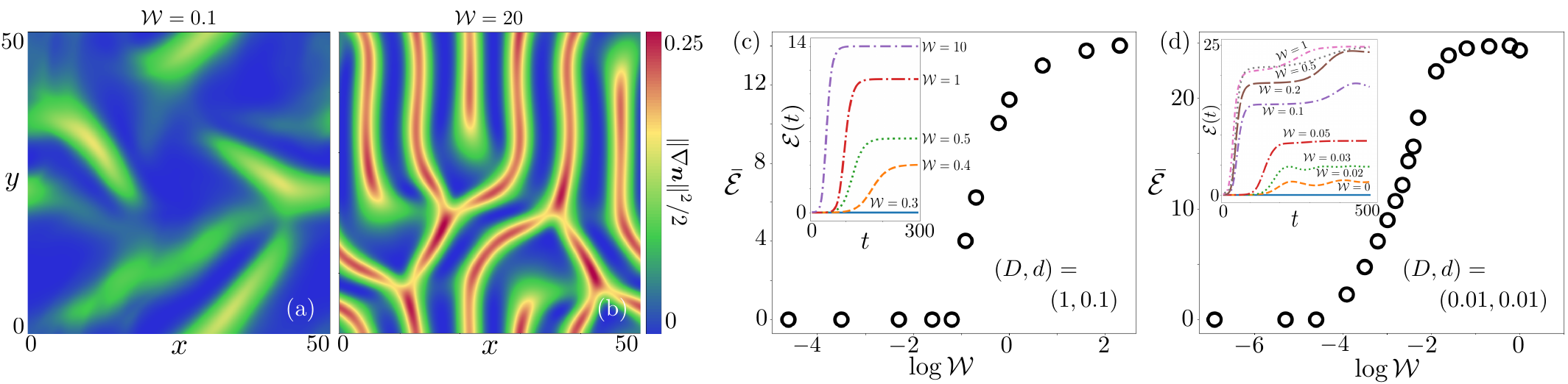}
\caption{(a) At large particle activity, $\Er=-70$, and small anchoring strength, $\W=0.1$, a chaotic flowing state emerges as in Newtonian fluids -- the state shown is highly unsteady. (b) At large anchoring strength, $\W=20$, the system settles into an arrested state, with high wavenumber distortions and pseudo-defects as in active nematics -- the state shown is a steady, flowing state. See Movie 7. (c) The mean elastic energy at late times, computed using the full kinetic theory for a range of anchoring strengths, $\W$, with $(\Er, \eta, \volfrac, \gamma) = (-8,1,0.02,0.1)$ fixed, and large diffusion constants, $(D,d) = (1,0.1)$. A critical transition is observed at $\mathcal{W}\approx 0.3$. Inset: transient evolution of the elastic energy, $\mathcal{E}(t)$. (d) Same as in (c), but for smaller diffusion constants, $(D,d) = (0.01,0.01)$. With smaller diffusion (e.g. smaller thermal energy), the anchoring strength needed to trigger a non-trivial state is also lower. See Movie 8.}
\label{Figure2}
\end{figure}

The anchoring strength selects the extent to which the system acts as an active suspension in a transversely isotropic medium \cite{hcsgcd18} ($\W\to 0$), or as an active nematic ($\W \to \infty$). Figure~\ref{Figure2}a,b (and Movie 7) show two states using the same parameters as in Fig.~\ref{Figure1}b, but with stronger activity, $\Er=-70$. In the first panel, the anchoring strength is small, $\W=0.1$, and the system undergoes chaotic dynamics as in a Newtonian fluid \cite{ss08}, while at large anchoring strength, $\W=20$, a steady (flowing) labyrinthine arrested state emerges \cite{ajc20,lajc24,pmis24}. 


Figure~\ref{Figure2}c shows the mean elastic energy for late times (long after the initial instability has occurred), defined by
\begin{gather}
    \bar{\mathcal{E}}:=\frac{1}{100}\int_{400}^{500}\int_0^\LL\int_0^\LL\frac{1}{2}|\nabla\Phi|^2\,dx\,dy\,dt.
\end{gather}
This energy is computed using the full kinetic theory as a function of the anchoring strength, revealing a phase transition at roughly $\W \approx 0.3$ for parameters $(\Er, \volfrac,\gamma,\eta, D, d) = (-8,0.02,0.1,1,1,0.1)$, the same used to create Fig.~\ref{Figure1}b. For these parameters the system relaxes to perfect alignment (equilibrium) below this value of $\W$, and settles into the first flowing arrested state (like that in Fig.~\ref{Figure1}b) for large $\W$.

More complex dynamics ensue when diffusion is reduced. Figure~\ref{Figure2}d shows $\bar{\mathcal{E}}$ again as a function of the anchoring strength, but instead using $(D, d) = (0.01,0.01)$. For small but finite anchoring strengths, $\W \in (0.01,0.1)$, the LC bulk elastic energy increases almost linearly with the anchoring strength. At $\W \approx 0.1$ and $0.15$, the spatial bands of large elastic energy break, split, and relax towards the first arrested state, like that in Fig.~\ref{Figure1}b, but at a 45 degree angle relative to the square periodic domain (see Movies 7, 8). For larger values of $\W$, the system relaxes to states similar to the second arrested state, as in Fig.~\ref{Figure1}c. Thus, with smaller diffusion, a smaller anchoring strength is needed to trigger an arrested state. Ultimately, then, the nature of the phase transition in $\W$, or even the question of whether there is a phase transition, depends on diffusion, a standard feature of both passive \cite{Bray02} and active \cite{ct15} suspensions.

\section{From the bend instability to exact arrested states} \label{sec:exact arrested states}

Moment equations can provide insight into stability criteria and the fully nonlinear arrested state. We consider a locally aligned particle distribution $\psi(\x, \p, t) \approx c(\x,t)\delta(\p-\m)$, with first and second orientational moments $\m:=\langle \p \rangle/c$ and $\D = \left<\p\p \right>/c \approx \m\m$. Using Eq.~\eqref{eqn:AMpsi} we find (with $D_t := \partial_t + \u\cdot \nabla$):
\begin{gather}
D_tc +\V  \nabla \cdot \left(c\m\right)  = D \nabla^2 c,\label{eqn:moment0} \\
D_t(c\m) + \V \nabla \cdot \left(c\m\m\right)= c\left(\I \m-\m \m \m \right) : \left(\nabla \u + \W\eta^{-1} \n \n\right) + D\nabla^2(c\m)-2dc\m, \label{eqn:moment1}
\end{gather}
along with Eq.~\eqref{eqn:LCdir} with $\lambda=1$. Motivated by Figs.~\ref{Figure1}b and \ref{Figure1}d, we seek solutions which only vary in $x$. Writing $c = c(x,t)$, $\m = (\cos{\Theta(x,t)}, \sin{\Theta(x,t)},0)$, $\n = (\cos{\nphi(x,t)}, \sin{\nphi(x,t)},0)$ and $\u = (0, v(x,t),0)$, Eqs.~\eqref{eqn:moment0}, \eqref{eqn:moment1} and \eqref{eqn:LCdir} reduce to:
\begin{gather}
    c_t + \V \left(\cos(\Theta) c_x-\sin(\Theta)\Theta_x c\right)=D c_{xx} ,\label{eqn:SA_c}\\ 
    \Theta_t + \V \cos(\Theta)\Theta_x= \cos^2(\Theta) v_x-\frac{\W}{2\eta } \sin[2(\Theta-\nphi)]+D\left(\Theta_{xx}+ \frac{2 c_x}{c}\Theta_x\right), \label{eqn:Theta}\\
    \nphi_t = \cos^2(\nphi) v_x + \frac{\W c}{2 \gamma} \sin[2(\Theta-\nphi)]+\frac{1}{\gamma}\nphi_{xx}. \label{eqn:phi}
\end{gather}
In the limit of small dimensionless rotational viscosity, $\gamma \to 0$, we find the particle orientation field dynamics 
\begin{gather}
    \partial_t\Theta_0 + \V  (\sin\Theta_0)_x=-\Er \volfrac \, c \cos ^3\Theta_0 \sin\Theta_0 + \frac{2D}{c} c_x \, (\Theta_0)_x +((\eta c)^{-1}+D) (\Theta_0)_{xx},
    \label{eqn:Theta0}
\end{gather}
 (see Appendix~\ref{appendix: arrested state calculation}). To study the first arrested state (Fig.~\ref{Figure1}b), we consider immotile particles, $\V =0$. By \eqref{eqn:SA_c}, the concentration relaxes to uniformity, $c=1$, and the dynamics are governed by \eqref{eqn:Theta0} alone. Defining $\xi=2\pi \LL^{-1} x$, at equilibrium, denoted by $\Theta_0(x,t) = \Teq(\xi)$, we find
\begin{equation}\label{eqn:Theta0_equilibrium}
\partial_{\xi\xi}\Teq +\Act\cos^3 \Teq \sin\Teq=0,
\end{equation}
where we have introduced an activity parameter
\begin{gather}\label{eq:Act}
    \Act= \frac{- \Er \volfrac \LL^2}{(2\pi)^2(D+\eta^{-1})},
\end{gather}
which incorporates not only the competition between active and passive elastic stresses (and particle volume fraction) represented by $\Er$, but also reduces the effective activity due to diffusivity. Note that $\Act>0$ for extensile stresses. A similar equation describes an active nematic steady state \cite{ajc20,lajc24}. 

Multiplying by $\Teq_\xi$ and integrating yields
\begin{equation}
    \frac{1}{2}\Teq_\xi^2 -\frac{\Act}{4} \cos^4 (\Teq)=C
\end{equation}
for some constant $C$. Choosing $\Teq(0)=\|\Teq\|_{\infty}:=\vartheta$ and $\Teq_\xi(0)=0$ (which we may do without loss of generality, by phase invariance), $C$ may be written in terms of $\vartheta$, yielding
\begin{equation}
    \Teq_\xi^2 =\frac{\Act}{2}\left(\cos^4 \Teq-\cos^4 \vartheta\right).
\end{equation}
We will focus on extensile particles, $\Act>0$. On the quarter domain $\xi\in[0,\pi/2$] we have $0 \leq \Teq(\xi)\leq \vartheta$, and we may choose without loss of generality that $\Teq_\xi<0$ and $\Teq(\pi/2)=0$ at the boundary. This selects the appropriate branch of the square root,
\begin{equation}\label{eqn:sqrtselection}
    \Teq_\xi =-\sqrt{\frac{\Act}{2}}\sqrt{\cos^4 \Teq-\cos^4 \vartheta},
\end{equation}
a separable equation. The trivial solution, $\Teq=0$ and $\vartheta=0$, is one solution. Other solutions around found by integration,
\begin{equation}\label{eqn:integration}
\int_\vartheta^{\Teq} \frac{d\Teq}{\sqrt{\cos^4 \Teq-\cos^4\vartheta}} =-\sqrt{\frac{\Act}{2}}\xi.
\end{equation}

The above integral yields a closed-form representation of $\Teq$ in terms of the maximal angle $\vartheta$,
\begin{equation}\label{eqn:Theta0_sln}
\begin{aligned}
\Teq(\xi)
=\cos^{-1}\left(\frac{\cos (\vartheta)}{\sqrt{1-2\,{\rm sn}^2\left( B_1 \xi ; B_2\right)}}\right),
\end{aligned}
\end{equation}
where $B_1=(\Act/8)^{1/2} \sin (2 \vartheta)$ and $B_2=2 \csc ^2\vartheta$. Here ${\rm sn}$ is a Jacobi elliptic function, which is sinusoidal if $B_2=0$ and a square wave if $B_2=1$. The maximum particle orientation angle, $\vartheta$, is found by setting $\Teq(\pi/2)=0$,
\begin{equation}\label{eq:constraint}
   \frac{-F(\vartheta)}{\cos^2\vartheta}=-\frac{\pi}{2}\sqrt{\frac{\Act}{2}},
\end{equation}
where $F(\vartheta) = \int_0^\vartheta \left(\cos^4\Theta/\cos^4\vartheta-1\right)^{-1/2}\,d\Theta= K\left(\sin ^2\vartheta/2\right)\cos\vartheta /\sqrt{2}$ and $K(\cdot)$ is the complete elliptic integral of the first kind. The maximum angle, $\vartheta$, is thus selected by the solution to 
\begin{gather}\label{eq:vartheta_eqn}
  \frac{2}{\pi}K\left(\frac{\sin ^2\vartheta}{2}\right)-\sqrt{\Act} \cos\vartheta=0.
\end{gather}   

%

Examining the limit as $\vartheta \to 0$, since $K(0) = \pi/2$, the critical value of $\Act$ for which a non-trivial solution appears is $\Act = 1$. Solutions begin to emerge when $\Act\geq 1$, increasing continuously from $\vartheta=0$. For values of $\Act$ just larger than 1, expanding around small $\vartheta=0$ to second order, we find
\begin{gather}\label{eqn:approx1}
    \vartheta \approx\sqrt{\frac{2(\sqrt{\Act}-1)}{\sqrt{\Act}+1/4}}.
\end{gather}
Expanding instead around $\vartheta=\pi/2$, we find
\begin{gather}\label{eqn:approx2}
    \vartheta \approx \frac{\pi}{2}-\frac{\Gamma\left(\frac{1}{4}\right)}{\sqrt{2\pi \Act} \, \Gamma\left(\frac{3}{4}\right)}\approx \frac{\pi}{2}-\frac{1.18}{\sqrt{\Act}}.
\end{gather}

\begin{figure}
    \centering
    \includegraphics[width=.7\textwidth]{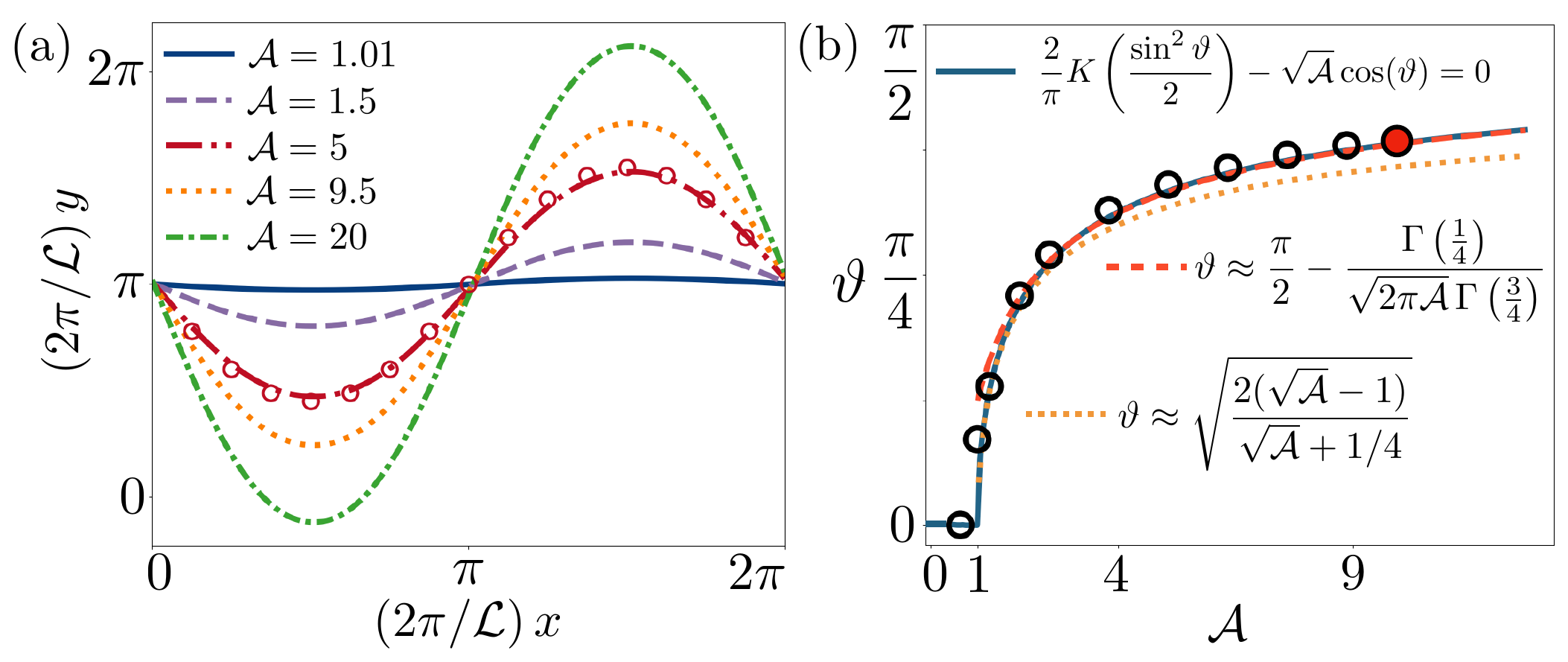}
    \caption{(a) Analytical director lines for a range of activity parameters, $\Act$. Symbols correspond to the case shown in Fig.~\ref{Figure1}b, obtained using the full kinetic theory, showing close agreement of the analytical approximation. (b) The maximum orientation angle at equilibrium, $\vartheta=\|\Teq\|_{\infty}$, as a function of $\Act$, at infinite anchoring strength. Solid, dashed, and dotted lines are from numerical solution of \eqref{eq:vartheta_eqn}, and approximations for $\Act \gg 1$, and for $\Act\approx 1$, respectively; symbols, meanwhile, are found by numerical simulation of the moment equations~\eqref{eqn:moment0}-\eqref{eqn:moment1}. The filled symbol is a second critical $\Act$ beyond which the first arrested state is unstable to transverse perturbations.}
    \label{Figure3}
\end{figure}

The director lines associated with \eqref{eqn:Theta0_sln} (integral curves, with $(2\pi y/\LL)=\int_0^\xi \tan\Teq(s)\,ds$) are shown in Fig.~\ref{Figure3}a for a range of activity parameters, $\Act$, showing good agreement with the computed results using the full kinetic theory. The symbols in Fig.~\ref{Figure3}a correspond to the numerically observed director line in Fig.~\ref{Figure1}b (where $\Act \approx 5$). The value of $\vartheta$ obtained from Eq.~\eqref{eq:vartheta_eqn} is plotted in Fig.~\ref{Figure3}b as a solid line for different values of $\Act$, along with symbols showing values computed using full 2D simulations of the moment equations, ~\eqref{eqn:moment0}-\eqref{eqn:moment1}, which are in close agreement. The asymptotic approximations for $\vartheta$ above are included for reference, for $\Act\approx 1$ and $\Act\gg 1$, revealing the initial bifurcation to the first arrested state at $\Act=1$. 

Although arrested solutions exist for all $\Act>1$, they become unstable to transverse perturbations beyond a critical activity strength. Simulations of the moment equations suggest that this transverse instability is triggered, using the same parameters as used in Fig.~\ref{Figure1}b at $\Act\approx 9.5$, indicated by a filled symbol in Fig.~\ref{Figure3}b.



\section{Motile particles surf on self-generated waves}\label{Traveling waves}
\begin{figure}
\includegraphics[width=0.9\textwidth]{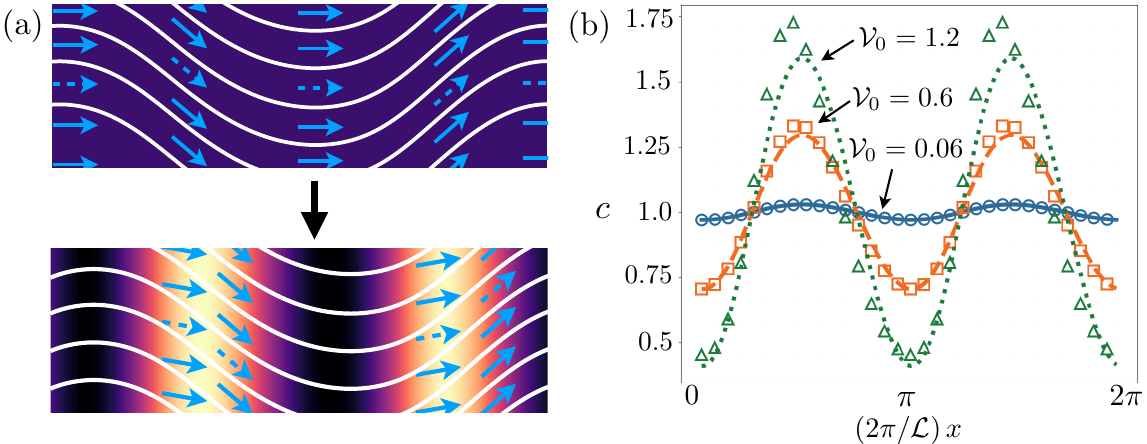}
\caption{\label{fig:Figure4} (a) Horizontal particle velocities vary in the arrested state due to a geometric effect. Motile particles in bent regions move horizontally with their swimming speed, while the swimming motion of particles elsewhere is redirected vertically. This leads to faster particle evacuation from regions with large LC bending, resulting in traveling bands of particles `surfing' ahead of the bent regions. But LC bending is generated collectively by the particles themselves, resulting in continuous wave propagation. (b) Theoretical concentration deviations (curves) for fixed $\Act = 2$ (the same activity parameter used in Fig.~\ref{Figure1}d) with dimensionless swimming speeds $\V  \in \{0.06, 0.6, 1.2\}$. Symbols are computed results using the moment equations.} 
\end{figure}

We now take up the question of the periodic traveling waves observed for motile particles ($\V >0$), as in Fig.~\ref{Figure1}d. The physical mechanism is geometric in nature: due to motility, particles with a larger horizontal velocity component encroach upon particles with a smaller such component, thus evacuating from regions with large LC bending and clustering in regions of small LC bending, as illustrated in Fig.~\ref{fig:Figure4}a. For intermediate activity parameter, $\Act$, motile particles tend to `surf' ahead of the bent LC that their active stress has created, drawing the bent state forward along with them. This is reminiscent of other systems in which particles interact with their own generated wave fields \cite{cpfb05} -- `pilot-wave' dynamics which have led to a wide range of hydrodynamic-quantum analogs \cite{Bush15,bo20} -- though here the generated wave is the undulatory background LC configuration, generated not by a single particle but by the collective effort of the entire suspension. 

The wave speed, $\omega$, may be estimated by studying \eqref{eqn:SA_c} and \eqref{eqn:Theta0} in a comoving frame, $\xi := 2\pi \LL^{-1} (x-\omega t)$, where $\omega$ is not yet known. Using that $(c,\Theta_0)=(1,\Teq)$ is a periodic solution with trivial wave speed $\omega=0$ when $\V =0$, we again pursue a regular perturbation expansion $c = 1 + \V  c^{(1)}(\xi) +  \V ^2 c^{(2)}(\xi) + \cdots$, $\Theta_0 = \Theta^{(0)}(\xi) + \V  \Theta^{(1)} (\xi) + \V ^2 \Theta^{(2)}(\xi) + \cdots$, and $\omega = \V  \omega^{(1)} + \V ^2 \omega^{(2)} + \cdots$. At leading order, we have that $\Theta^{(0)}=\Teq$, the steady solution found for $\V  = 0$, given in \eqref{eqn:Theta0_sln}. At $O(\V)$, we obtain a system of equations for $c^{(1)}$ and $\Theta^{(1)}$ in terms of $\Teq$,
\begin{gather}
    \partial_{\xi\xi}c^{(1)} = \frac{\LL}{2\pi D}\partial_\xi \cos\Teq, \label{eqn:TW_c1}\\
    \partial_{\xi\xi}\Theta^{(1)} = - \Act \left(\cos^4\Teq- 3 \cos^2\Teq \sin^2\Teq\right) \Theta^{(1)} + f(\xi), \label{eqn:TW_Theta1}
\end{gather}
where 
\begin{gather}
    f(\xi)
    = \left(\frac{2+\eta D}{1+\eta D}\right)c^{(1)}\partial_{\xi\xi}\Teq+\frac{\eta}{2\pi(1+\eta D)}\left(\LL(\cos\Teq-\omega^{(1)})-4\pi D c^{(1)}_\xi \right)\partial_\xi\Teq. \label{eqn:forcing}
\end{gather}

Integrating \eqref{eqn:TW_c1}, we obtain
\begin{gather}
    c^{(1)} = \frac{\LL}{2\pi D} \int_0^{\xi} \cos\Teq(s)\, ds +  k_1 \xi + k_2, \label{eqn:c1_expr}\\
    k_1 = - \frac{\LL}{(2 \pi)^2 D} \int_0^{2 \pi} \cos\Teq(s) \, ds, \quad k_2 = \pi k_1 + \frac{\LL}{(2 \pi)^2 D} \int_0^{2\pi} s\cos\Teq(s) \, ds,
\end{gather}
where $k_1$ and $k_2$ are determined, respectively, by requiring that $c^{(1)}_\xi$ is periodic and $c^{(1)}$ is mean zero. A comparison of the solution with numerical results is shown in Fig.~\ref{fig:Figure4}b, showing good agreement across a range of swimming speeds $\V$. For small $\V$, we find that the traveling wave speed is given to leading order by 
\begin{gather}
    \omega \sim \V \omega^{(1)} = \V \left[\frac{1}{2} \left(5 I_1-3 \frac{I_2}{I_3}\right)+\frac{1}{\eta D}\left(I_1-\frac{I_2}{I_3}\right) \right],\label{eqn:TW_omega1}
\end{gather}
where
\begin{gather}
I_1 = \frac{1}{2 \pi}\int_0^{2 \pi} \cos(\Teq) \, d\xi,\,\,\,\,
I_2 = \int_0^{2 \pi} \cos (\Teq) \Teq_{\xi}^2 \, d\xi,\,\,\,\,
I_3 = \int_0^{2 \pi} \Teq_{\xi}^2 \, d\xi,
\end{gather}
(see Appendix~\ref{appendix: traveling wave speed calculation}). The expression \eqref{eqn:TW_omega1} is exact for the linearized equations \eqref{eqn:TW_c1}-\eqref{eqn:TW_Theta1}, so any discrepancy from the behavior of equations \eqref{eqn:SA_c} and \eqref{eqn:Theta0} is due to linearizing about the $\V =0$ steady state. 
The magnitude of this error is $O(|\V c^{(1)}|^2,|\V \Theta^{(1)}|^2)$ as $\V \to 0$, which can be approximated as $O(\V^2\LL^2/D^2,\V^2\LL^2\Act^2/D^2)$ according to the Eqs.~\eqref{eqn:Theta0_equilibrium}, \eqref{eqn:sqrtselection}, and \eqref{eqn:c1_expr}, suggesting that \eqref{eqn:TW_omega1} should give a reliable prediction at small swimming Péclet number, $\Pe:=\V/D = (a V_0^*)/D^* \ll 1$, and intermediate activity, $\Act$ (larger than unity but not dramatically so).

\begin{figure}
    \centering
    \includegraphics[width=0.7\textwidth]{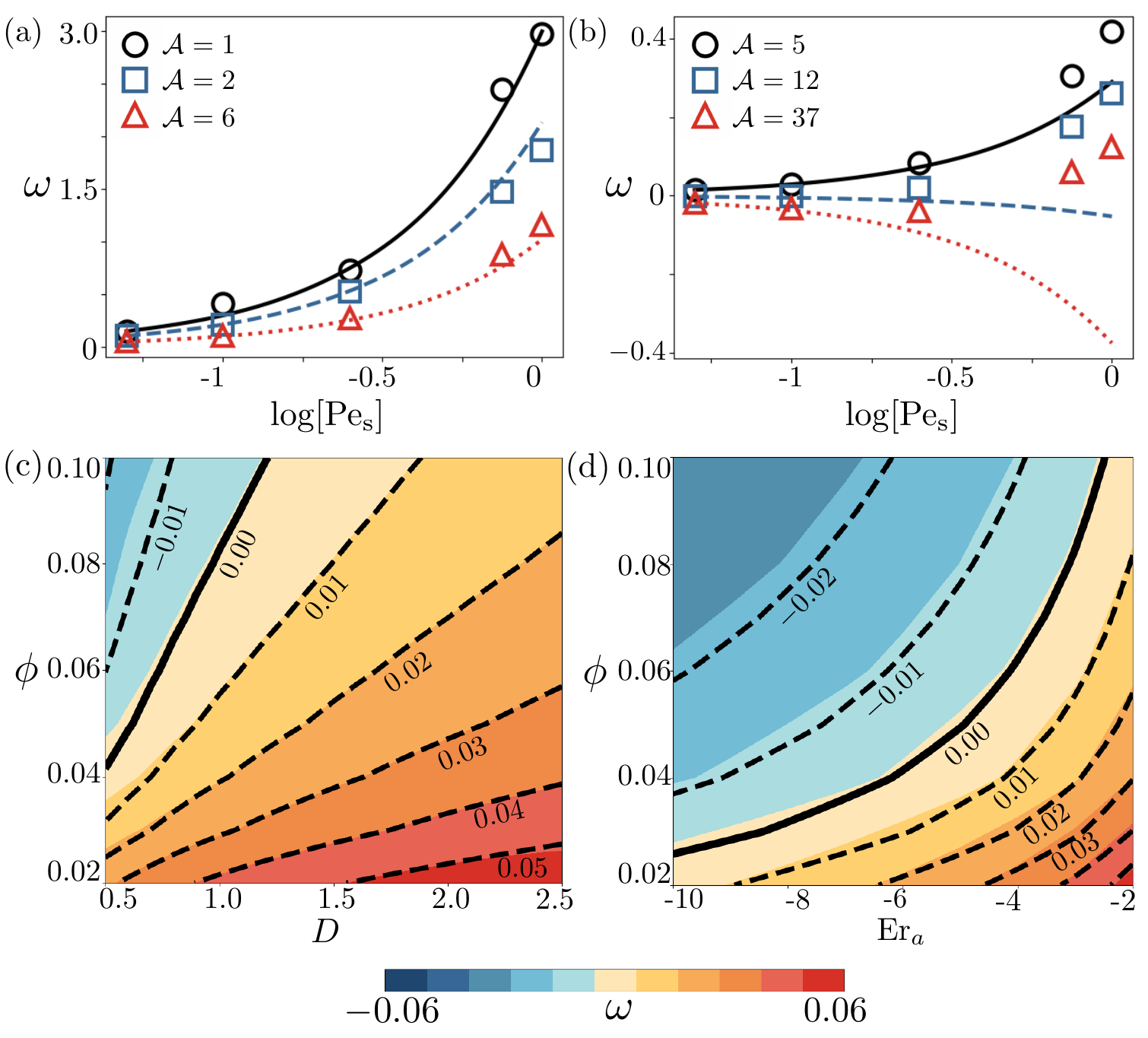}
    \caption{(a,b) The wave speed $\omega$ vs.~$\log[\Pe]$, where $\Pe=aV_0^*/D^* \in [0.05, 1.0]$ is the swimming Péclet number, for activity parameters $\Act \in \{1,2,6\}$ (a), and $\Act \in \{5,12,37\}$ (b). Values computed using the reduced 1D system~\eqref{eqn:SA_c} and \eqref{eqn:Theta0} are shown as symbols, and the curves with matching colors show the analytical value from \eqref{eqn:TW_omega1}. The analytical prediction is accurate so long as the activity parameter is not too large. (c) Filled contours of wave speeds are generated using numerical simulations of \eqref{eqn:SA_c} and \eqref{eqn:Theta0} for fixed $\V = 0.06$, are compared with the theoretical expression~\eqref{eqn:TW_omega1} (shown as dashed curves) for a range of volume fractions $\volfrac$ and dimensionless diffusivities $D$. Increasing $\volfrac$ or decreasing $D$ both increase the activity parameter, $\Act$, which increases the maximum bending angle in the LC, which inhibits lateral motion and reduces the wave speed. For extreme values, the propagating wave even reverses direction. (d) The same comparison, but using the active Ericksen number, $\Er$. Similarly, larger negative $\Er$ corresponds to larger $\Act$, thus larger $\vartheta$, and smaller wave speeds. Here again, direction reversal is possible (the wave passes in the opposite direction as the mean particle motion for $\omega<0$).
    }
    \label{fig:Figure5}
\end{figure}

To further explore the accuracy of \eqref{eqn:TW_omega1}, Figs.~\ref{fig:Figure5}a,b show the predicted wave speeds as solid curves across a range of swimming Péclet numbers, $\Pe=a V_0^*/D^*$, for six different activity coefficients, $\Act$. Symbols represent the wave speeds computed using \eqref{eqn:SA_c} and \eqref{eqn:Theta0}\footnote{An additional comparison to values computed using the 2D moment equations is included in the Supplementary Materials, showing equally good agreement}. The analytical approximations are in good agreement for intermediate $\Act$, particularly as $\V /D \to 0$ as expected. For larger swimming speeds the physical concentration bands become sharper, and the traveling wave moves faster, since the differential horizontal swimming component where the LC is most and least bent is more pronounced. Larger activity, meanwhile, serves to reduce the wave speed, since the traveling arrested state has a larger bend, reducing the mean horizontal directed motion. The analytical estimates begin to suffer at large activity parameter and large swimming Péclet number, but the analysis does capture the wave direction reversal for large $\Act$ and small $\Pe$. For large $\Act$, Fig.~\ref{fig:Figure5}b indicates that the traveling wave undergoes yet another direction reversal for large $\Pe$, so that the wave once again travels in the same direction as the active particle motion. 

To better understand the role of particle volume fraction on the wave speed, the system is more clearly described in terms of $\volfrac$ and an independent measure of active stress, $\Er$ (since $\Act$ couples the two). Figure~\ref{fig:Figure5}c shows contours of the wave speed as a function of the volume fraction and the relative diffusion ($D=\mu D^*/K$) from numerical simulation of \eqref{eqn:SA_c} and \eqref{eqn:Theta0} with $\V =0.06$ and $\Er=-3$ fixed, and those from the theoretical expression in \eqref{eqn:TW_omega1} shown as dashed curves. Increasing $D$ reduces the maximum orientation angle in the traveling arrested state, $\vartheta$ (since $\Act$ is inversely proportional to $D$ in \eqref{eq:Act}), which increases the wave speed. Increasing the volume fraction, $\volfrac$, has the opposite effect - it increases the maximum orientation angle, $\vartheta$, by increasing the relative activity, $\Act$, and thus impedes lateral motion and decreases the wave speed. 

Finally, Fig.~\ref{fig:Figure5}d shows a different cross-section through parameter space, this time with $\V=0.06$ and $D=1$ fixed, but varying the active Ericksen number. Ultimately, for small $\V$ at least, the wave speed is again dictated by the maximum orientation angle in the traveling arrested state. Hence, particles with larger negative $\Er$ (still for extensile-stress-generating particles) act to increase $\vartheta$, thus diminishing the wave speed, just as increasing the particle volume fraction does again in this case. Direction reversal to retrograde wave propagation at larger swimming Péclet number is captured here as well. The concentration wave passes opposite the direction of swimmer motion for sufficiently large $\volfrac$, though this is counteracted by large diffusivity, or small active stress. 

Finally, as a simple approximation, with $\Act > 1$ but not too large, taking $\Teq\approx \vartheta \cos\xi$ (see Fig.~\ref{Figure3}b), we find 
\begin{gather}
    \omega \approx  \frac{9\V}{8 \sqrt{\Act}}.
\end{gather}
This estimate is comparable to the values reported in Fig.~\ref{fig:Figure5}b. The wave speed is naturally enhanced by faster swimming, but is reduced by larger activity or particle volume fraction, or smaller LC elasticity, since these contribute to larger LC deformations (larger $\vartheta$ via $\Act$), again hindering particle transport. This approximation does not capture direction reversal, however, which only appears for larger $\Act$.

\section{Conclusion} 

We have shown that the bend instability of a suspension of active extensile particles may be tamed by elastic stresses in a surrounding anisotropic, viscoelastic fluid, leading to arrested, flowing states beyond a critical activity parameter, $\Act\geq 1$. This activity parameter is proportional to the active stress, particle volume fraction, and system size, and is inversely proportional to particle diffusivity and LC elasticity. The degree of LC deformation in the arrested state depends on this activity parameter, which can result in additional transverse instabilities and higher wavenumber textures.

When extensile-stress-generating particles are motile, traveling concentration waves ensue. The horizontal component of swimming velocity is largest in regions of largest LC bending, leading to evacuation of the bent regions and concentration bands of `surfing' particles just ahead of them. The LC configuration, which is generated by the particles, then also translates, resulting in a steady, propagating arrested state. The particle swimming speed and the degree of LC bending dictate the speed of wave propagation, so the wave speed decreases with increased activity parameter $\Act$; e.g., with increased active particle stress, or decreased LC elasticity, which both increase LC bending. More exotic dynamics are revealed with larger particle activity, including a dramatic, periodic thrashing mode.

The model presented here represents a continuous interpolation from active suspension theory in Newtonian fluids, to active nematics, tuned by the anchoring strength, which features a phase transition to arrested states at a critical anchoring strength. The nature of the phase transition was observed to depend on the diffusion, with smaller diffusion constants requiring smaller anchoring strengths to provide sufficient particle-LC alignment and coherence to trigger a buckling instability. 

It is imaginable that the use of Ericksen-Leslie theory could suppress the formation of topological defects at high active Ericksen number, high particle concentration, and large anchoring strength. However, since solutions were recovered in all cases considered without the development of even nearly singular behavior, we expect the results described in the studied regime to carry over to a higher order treatment of the nematic LC.

Future work is needed; we did not explore the role of a number of parameters (e.g.~the tumbling parameter, $\lambda$). Nor did we include anisotropic viscous response 
 \cite{hcsgcd18}, particle deformability \cite{cs24b}, or elastic interactions between active particles through the LC, which has been considered in Refs.~\cite{szla15,ztgsal17}, all of which may be important in some manifestations of this system. Nevertheless, the analytical descriptions of the arrested states and traveling wave speeds can provide a benchmark for examining more complex states which emerge at yet higher particle concentrations, which should be useful given the large number of parameters needed to characterize the systems of interest.

\textit{Acknowledgements.} We are grateful for helpful discussions with Thomas G. J. Chandler and Ido Lavi, and suggestions by anonymous reviewers. Support for this research was provided by the VCRGE with funding from the Wisconsin Alumni Research Foundation. LO acknowledges support by the NSF (DMS-2406003). 

\appendix

\section{Arrested states at large anchoring strength}\label{appendix: arrested state calculation}

Here we carry out the calculation leading to the leading order dynamics for the particle orientation field in \S~\ref{sec:exact arrested states}.
From Eq.~\eqref{eqn:Stokes}, we have
\begin{gather}
    -p_x + \partial_x\left[-(\nphi_x)^2+\frac{\lambda}{2}\left(\nphi_{xx} + \frac{\W c}{2}\sin[2(\Theta-\nphi)]\right)\sin(2\nphi)+\Er \volfrac c \cos^2{\Theta}\right] = 0, \\
    v_{xx} + \partial_x\left[(-1-\lambda\cos(2\nphi))\left(\frac{1}{2}\nphi_{xx}+ \frac{\W c}{4}\sin[2(\Theta-\nphi)]\right)+\frac{\Er \volfrac c}{2}\sin{2\Theta}\right] = 0.
\end{gather}
Using periodicity, the fluid velocity and pressure satisfy
\begin{gather}
    p(x) = -(\nphi_x)^2+\frac{\lambda}{2}\left(\nphi_{xx} + \frac{\W c}{2}\sin[2(\Theta-\nphi)]\right)\sin(2\nphi)+\Er \volfrac c \cos^2{\Theta}, \label{eqn:p}\\
    v_x = (\lambda\cos(2\nphi)+1)\left(\frac{1}{2}\nphi_{xx}+ \frac{\W c}{4}\sin[2(\Theta-\nphi)]\right)-\frac{\Er \volfrac c}{2}\sin{2\Theta} \label{eqn:v}.
\end{gather}

For large anchoring strength, $\W \gg 1$, we pursue a regular perturbation expansion, 
\begin{gather}
\nphi = \nphi_0(x,t) +  \W^{-1}  \nphi_1(x,t) +  \W^{-2} \nphi_2(x,t) + \cdots, \\
 \Theta = \Theta_0(x,t) + \W^{-1} \Theta_1(x,t) + \W^{-2}  \Theta_2(x,t) + \cdots.
\end{gather}

 At leading order, from Eq.~\eqref{eqn:Theta} we find $\nphi_0=\Theta_0$. Inserting this relation into the velocity gradient and expanding, we find that
\begin{gather}
    v_x =-\frac{\Er \volfrac c}{2} \sin (2 \Theta_0)+ \frac{1}{2}(\lambda  \cos (2 \Theta_0)+1)\left( \partial_{xx}\Theta_0+c (\Theta_1-\nphi_1)\right)+O\left(\frac{1}{\W}\right)
\end{gather}
as $\W \to \infty$. Eq.~\eqref{eqn:Theta} then provides an equation for the dynamics of $\Theta_0$ (or equivalently for $\nphi_0$) which, through the velocity field, depends on the difference of the active and passive fields at the next order: 

\begin{gather}
\partial_t\Theta_0+\V\cos(\Theta_0)\partial_x\Theta_0=-\Er \volfrac c\cos^3\Theta_0\sin\Theta_0 +\left(c \, g(\Theta_0) -\frac{1}{\eta}\right) (\Theta_1-\nphi_1)
+(g(\Theta_0)+D)\partial_{xx}\Theta_0 + \frac{2D c_x}{c} \partial_{x}\Theta_0, \label{eqn:nextdiff1}
\end{gather}
where $g(\Theta_0) = \frac{1}{2}\cos^2(\Theta_0)\left(\lambda\cos(2\Theta_0)+1\right)$. However, from Eq.~\eqref{eqn:phi} (replacing $\nphi_0$ with $\Theta_0$), we also have that 
\begin{gather}
    \partial_t \Theta_0 =-\Er \volfrac c \cos^3\Theta_0 \sin \Theta_0+  \left(g(\Theta_0)+\frac{1}{\gamma }\right)\left[c (\Theta_1-\nphi_1)+\partial_{xx} \Theta_0\right].
    \label{eqn:nextdiff2}
\end{gather}
Combining Eqs.~\eqref{eqn:nextdiff1} and \eqref{eqn:nextdiff2}, we find
\begin{gather}
    \nphi_1(x) -  \Theta_1(x)=\frac{1}{c(c+\gamma/\eta)}\left(\gamma (  \V c\cos(\Theta_0) - 2D c_x)\partial_x \Theta_0+(1-\gamma  D)c \, \partial_{xx}\Theta_0\right),
\end{gather}
and so obtain an equation for $\partial_t\Theta_0$ depending only on $\Theta_0$ and $c$:
\begin{gather}
    \partial_t\Theta_0=-\Er \volfrac c \cos ^3\Theta_0 \sin \Theta_0+ \frac{ g(\Theta_0)+\gamma^{-1}}{1+ c \eta \gamma^{-1}}\left(-\eta (\V c \cos(\Theta_0)-2D c_x)\partial_x \Theta_0+(1+ \eta D  c)\partial_{xx}\Theta_0\right). \label{eqn:Theta0withgamma}
\end{gather}

Taking the limit of small dimensionless rotational viscosity, $\gamma \to 0$, we arrive at Eq.~\ref{eqn:Theta0}.

\section{Traveling arrested states - wave speed}\label{appendix: traveling wave speed calculation}

Here we solve for the traveling wave speed as discussed in \S~\ref{Traveling waves}. With $c^{(1)}(\xi)$ determined explicitly in terms of $\Teq$ in \eqref{eqn:c1_expr}, the term $f(\xi)$ in Eq.~\eqref{eqn:TW_Theta1} may be considered as an inhomogeneous forcing term, suggesting a variation of parameters approach. First note that $\Teq_\xi$ is a solution to the homogeneous version of \eqref{eqn:TW_Theta1} ($f=0$), which can be seen as follows. Using that $\Teq_{\xi\xi} = -\Act\cos^3\Teq\sin\Teq$, we have
\begin{gather}
     - \Act \left(\cos^4\Teq- 3 \cos^2\Teq \sin^2\Teq\right) \Teq_\xi 
     = -\Act \partial_{\Teq} \left(\cos^3\Teq \sin\Teq\right)\Teq_\xi= \partial_{\xi\xi\xi}\Teq.
\end{gather}
Since we know one solution $\Teq_\xi$ of the homogeneous second order equation, we can determine the other solution, $Y$, using the Wronskian $W_r(\xi) = \Teq_{\xi} Y_{\xi} - \Teq_{\xi \xi} Y$. In particular, we see that $(W_r)_{\xi} = 0$, so $W_r$ is a constant which we may take to be 1. Then $Y$ satisfies 
\begin{equation}
    Y_{\xi} - \left(\Teq_{\xi}\right)^{-1}\Teq_{\xi \xi} Y = \left(\Teq_{\xi}\right)^{-1},
\end{equation}
which can be solved exactly to yield 
\begin{equation}
    Y(\xi) = \Teq_{\xi} \int_0^{\xi}(\Teq_{\xi}(s))^{-2}\, ds.
\end{equation}
Note that $Y(\xi)$ is not $2\pi$-periodic. 
Using variation of parameters, the solution to the inhomogeneous version of Eq.~\eqref{eqn:TW_Theta1} with $f$ as in \eqref{eqn:forcing} is given by 
\begin{gather}
    \Theta^{(1)}(\xi) = G_1(\xi) \Teq_{\xi}(\xi) + G_2(\xi) Y(\xi), \label{eqn:TW_Theta1_sol}\\
    G_1(\xi) = G_1(0) - \int_0^\xi Y(s) f(s) \, ds , \quad G_2(\xi) = G_2(0) + \int_0^\xi \Teq_{\xi}(s) f(s) \, ds,
\end{gather}
where $(G_1)_{\xi} \Teq_{\xi} + (G_2)_{\xi} Y(\xi) = 0$. The periodicity of $\Teq$, and the non-periodicity of $Y(\xi)$, demand that
\begin{gather}\label{eqn:cond3}
    G_2(2 \pi) -G_2(0) = \int_0^{2 \pi} \Teq_{\xi} f \, d\xi = 0.
\end{gather}

Inserting \eqref{eqn:forcing} into \eqref{eqn:cond3}, and rearranging, we require
\begin{equation}
\begin{aligned}
     \omega^{(1)} \int_0^{2 \pi} \Teq_{\xi}^2 \, d\xi
    &= \int_0^{2 \pi} \cos(\Teq) \Teq_{\xi}^2 \, d\xi-\frac{\pi(2+5\eta D)}{\eta \LL}\int_0^{2 \pi}c_\xi^{(1)}\Teq_{\xi}^2\,d\xi.
\end{aligned}
\end{equation}
Then, using the expression~\eqref{eqn:c1_expr} for $c^{(1)}_\xi$, we obtain  
\begin{align}
    \omega^{(1)} \int_0^{2 \pi}\Teq_{\xi}^2\,d\xi 
    & = \int_0^{2 \pi} \cos(\Teq) \Teq_{\xi}^2\, d\xi - \frac{2+5 \eta D}{2 \eta D} \int_0^{2 \pi}\left(\cos (\Teq) + \frac{2\pi D}{\LL} k_1\right) \Teq_{\xi}^2\,d\xi \\
    &=\frac{1}{2} (5 I_1 I_3-3 I_2)+\frac{1}{\eta D}\left(I_1 I_3-I_2\right),
\end{align} 
where we have defined
\begin{gather}
I_1 = \frac{1}{2 \pi}\int_0^{2 \pi} \cos(\Teq) \, d\xi,\,\,\,\,
I_2 = \int_0^{2 \pi} \cos (\Teq) \Teq_{\xi}^2 \, d\xi,\,\,\,\,
I_3 = \int_0^{2 \pi} \Teq_{\xi}^2 \, d\xi.
\end{gather}
This is the result provided in \eqref{eqn:TW_omega1}.

 The expression \eqref{eqn:TW_omega1} is exact for the linearized equations \eqref{eqn:TW_c1}-\eqref{eqn:TW_Theta1}, so any discrepancy from the behavior of equations \eqref{eqn:SA_c} and \eqref{eqn:Theta0} is due to linearizing about the $\V =0$ steady state. This error is $O(|\V c^{(1)}|^2,|\V \Theta^{(1)}|^2)$. 
Immediately from the expression~\eqref{eqn:c1_expr} for $c^{(1)}$, we see that $|c^{(1)}|\sim \LL D^{-1}$.
 %
 %
For $|\Theta^{(1)}|$, we first note that $\Teq_{\xi}\sim\sqrt{\Act}$, by Eq.~\eqref{eqn:sqrtselection}, while $\Teq_{\xi\xi}\sim \Act$, by Eq.~\eqref{eqn:Theta0_equilibrium}. Using the scalings $c^{(1)}\sim \LL D^{-1}$ and $\omega^{(1)}\sim \sqrt{\Act}^{-1}$, we have that the forcing term $f$ from Eq.~\eqref{eqn:forcing} roughly satisfies $f\sim \sqrt{\Act}\LL\eta(1+\eta D)^{-1}+\Act\LL D^{-1} \lesssim \Act\LL D^{-1}$ for $\Act\ge 1$. From the form~\eqref{eqn:TW_Theta1_sol} of $\Theta^{(1)}$, we then have $|\Theta^{(1)}|\sim \Act\LL D^{-1}$. The error in the linearized traveling wave speed $\V\omega^{(1)}$ is thus of size $O(\V^2\LL^2/D^2,\V^2\LL^2\Act^2/D^2)$.

\section{Glossary}

A glossary of terms is included in Tables~\ref{table1}-\ref{table3}. Though not independent of the groups in Table~\ref{table3}, we also define the swimming Péclet number, $\Pe:=a V_0^*/D^*$, and the active Péclet number, ${\rm Pe}_a:=\sigma a^2/(\mu v D^*)$.

\begin{table}\label{table1}
\caption{Physical quantities (particles)}
\centering
\begin{tabular}{c l} 
\hline\hline 
\\
$N$ & Number of particles\vspace{.3cm}\\
$a, b$ & Particle semi-major and semi-minor axis lengths\vspace{.3cm}\\
$v$ & Particle volume\vspace{.3cm}\\
$S$ & Particle surface area\vspace{.3cm}\\
$V_0^*$ & Particle swimming speed \vspace{.3cm}\\
$\sigma$ & Force dipole strength, $\sigma=2af$ with $f$ a force\vspace{.3cm}\\
$\eta$ & Rotational drag coefficient\vspace{.3cm}\\
$D^*$ & Translational diffusion constant\vspace{.3cm}\\
$d^*$ & Rotational diffusion constant\vspace{.3cm}\\
$L$ & Linear domain dimension\vspace{.3cm}\\
\hline
\end{tabular}
\end{table}

\begin{table}\label{table2}
\caption{Physical quantities (LC)}
\centering
\begin{tabular}{c l} 
\hline\hline 
\\
$K$ & Frank elastic constant\vspace{.3cm}\\
$W$ & Anchoring strength\vspace{.3cm}\\
$\mu$ & Viscosity\vspace{.3cm}\\
$\gamma^*$  & Rotational viscosity\vspace{.3cm}\\
\hline
\end{tabular}
\end{table}

\begin{table}\label{table3}
\caption{Dimensionless groups}
\centering
\begin{tabular}{c l} 
\hline\hline 
\\
$\LL = \displaystyle\frac{L}{a}$ & Relative system size\vspace{.3cm}\\
$\volfrac = \displaystyle\frac{N v}{L^3}$ & Particle volume fraction\vspace{.3cm}\\
$\Er = \displaystyle\frac{\sigma a^2}{v K}$ & Active Ericksen number\vspace{.3cm}\\
$\W =  \displaystyle\frac{SW}{aK}$ & Relative anchoring strength,\vspace{.3cm}\\
$\V = \displaystyle\frac{\mu a V_0^*}{K}$  & Relative swimming speed\vspace{.3cm}\\
$D =  \displaystyle\frac{\mu D^*}{K}$ & Relative translational diffusivity\vspace{.3cm}\\
$d = \displaystyle \frac{\mu a^2 d^*}{K}$ & 
Relative rotational diffusivity\vspace{.3cm}\\
$\gamma = \displaystyle \frac{\gamma^*}{\mu}$  & Relative rotational viscosity\vspace{.3cm}\\
$\lambda$ & LC tumbling parameter\vspace{.3cm}\\
\hline
\end{tabular}
\end{table}

\bibliographystyle{apsrev4-2}
\bibliography{BigBib}

\newpage

\section*{Supplementary Information}

We will use an Ericksen-Leslie description of the liquid crystal (LC) \citep{ll86,dp93}, which is assumed to be deep in the nematic phase. The local molecular orientation is denoted by $\n$. Director fields confined to two dimensions will be written in terms of a single angle field $\nphi$, or $\n = (\cos{\nphi},\sin{\nphi}, 0)$. In the one-constant approximation, the elastic energy density is $(K/2) \|\nabla' \n \|^2=(K/2a^2)\|\nabla \n \|^2$, with $K$ the Frank elastic constant, $\nabla'$ and $\nabla$ the del operators with respect to the dimensional and dimensionless positions $a\x$ and $\x$, respectively, and the norm above is the Frobenius norm.

We now consider the effect of introducing $N$ identically shaped active particles into the LC. The $j^{th}$ active particle position is written as $a\x_0^{(j)}$ and its orientations as $\p^{(j)}$. The boundary conditions are assumed to be finite-strength (`weak') tangential anchoring conditions with anchoring strength $W$. $K$ and $W$ have units of force and force per length, respectively. We assume that the bodies are large compared to the molecular constituents of the liquid crystal, but small compared to the length over which $\n$ is varying (a Type-VI system in the language of Ref.~\cite{su23}). We can then incorporate the associated moment into the volumetric energy density consistent with the Rapini-Papoular \cite{rp69,bd86} surface anchoring approximation. Denoting the energy density as $(K/a^2)\mathcal{F}$, we write
\begin{gather}\label{eqn:F1}
    \mathcal{F} = \frac{1}{2}\|\nabla  \n \|^2 + \left(\frac{a^2}{K}\right)\frac{SW}{2a^3 }\sum_{j=1}^N\delta(\x-\x_0^{(j)})\left(1-(\n \cdot \p^{(j)})^2\right).
\end{gather}
Here $S$ is the surface area of an individual particle; $S\sim \pi^2 ab(1+O((b/a)^2))$ for slender, rod-like particles. We have also defined a Dirac delta function on the dimensional spatial variables, $\delta^*(a\x)$, such that 
\begin{gather}
    \int_{\mathcal{D}^*}\delta^*(a\x-a\x_0)\,(a^3 dV)=1,
\end{gather}
for any point $a\x_0\in \mathcal{D}^*$, and a dimensionless delta function $\delta(\x-\x_0):=a^3\delta^*(a\x-a\x_0)$, so that 
\begin{gather}
    \int_{\mathcal{D}}\delta(\x-\x_0)\,dV=1.
\end{gather}

If the LC and active particle directions are confined to 2D, writing $\p^{(j)} = (\cos{\theta_j}, \sin{\theta_j},0)$, Eq.~\eqref{eqn:F1} is equivalent to
\begin{equation}
    \mathcal{F} = \frac{1}{2} |\nabla \nphi |^2 + \frac{\W}{2 }\sum_{j=1}^N \sin^2(\theta_j-\nphi)\delta(\x-\x_0^{(j)}),
\end{equation}
where we have introduced the dimensionless anchoring strength
\begin{gather}
   \W = \frac{SW}{aK}.
\end{gather}

Defining $\h := (\I -\n\n)\cdot \H$, where $\H = -\delta \mathcal{F}/\delta \n$ is the (dimensionless) LC molecular field \cite{dp93}, we have 
\begin{equation}
    \h = (\I -\n\n)\cdot\left(\nabla^2 \n+\W\sum_{j=1}^N (\n\cdot\p^{(j)})\p^{(j)}\delta(\x-\x_0^{(j)})\right).
\end{equation}
In two dimensions, we write $\h = h\n^{\perp}$, with $\n^{\perp} = (-\sin{\nphi}, \cos{\nphi}, 0)$, so that 
\begin{equation}
    h = \nabla^2 \nphi + \frac{\W}{2} \sum_{j=1}^N \sin(2(\theta_j-\nphi))\delta(\x-\x_0^{(j)}).
\end{equation}
Note that $h=0$ at equilibrium. For a continuum of active particles, in terms of $\psi^*$,
\begin{equation}
    \h = (\I -\n\n)\cdot \left(\nabla^2 \n + \W \int_{\mathbb{S}^{2}} \psi^*(\x^*, \p, t^*) (\n\cdot\p)\p \,d\p \right) = (\I -\n\n)\cdot \left( \nabla^2 \n + \W\n\cdot \left<\p\p\right> \right).
\end{equation}
In 2D, with $\h = h\n^{\perp}$, we have
\begin{equation}
    h = \nabla^2\nphi + \W\n^{\perp}\n : \left<\p\p \right>= \nabla^2\nphi + \W \n^{\perp}\n : \left<\p\p \right>,
\end{equation}
where $\boldsymbol{A}:\boldsymbol{B} = A_{ij}B_{ij}$, and we recall that $\langle \p\p \rangle$ has units of inverse volume. We write the stress corresponding to the elastic free energy as $\bsigma^*_r(a\x,T t):=(K/a^2)\bsigma_r(\x,t)$, where
\begin{equation}\label{eqn:sigmar}
    \bsigma_r(\x,t) = -\nabla \n \cdot \nabla \n^T - \frac{\lambda}{2}(\h \n + \n \h )+\frac{1}{2}(\h \n - \n \h).
\end{equation}
Here $\lambda$ is called the tumbling parameter (see Landau \& Lifschitz \cite[Ch.~6]{ll86}). 

We now move on to dynamics. At this point we are motivated select the timescale, $T=\mu a^2/K$, and we define the fluid velocity field $\u^*(a\x,Tt):=K(\mu a)^{-1}\u(\x,t)$, where $\mu$ is the solvent viscosity. The deviatoric viscous stress is written as $\bsigma_v^*(a\x,Tt):=(K/a^2)\bsigma_v(\x,t)$, where
\begin{equation}
    \bsigma_v(\x,t) = 2\mathbf{E} + \mu_1'\n \n(\n\cdot\mathbf{E}\cdot\n) + \mu_2'(\n\mathbf{E}\cdot\n+\n\cdot\mathbf{E}\n).
\end{equation}
Here $\mathbf{E} = (\nabla\u +\nabla \u^T)/2$ is the (dimensionless) symmetric rate of strain tensor, and $\mu_1'$ and $\mu_2'$ are dimensionless viscosity coefficients (which we take to be zero in the main text). Note that the $\mu_1'$ term absorbs an additional component associated with $\n\cdot \H$ , which appears in the active nematic models in, e.g., Ref.~\cite{lajc24}.

The particles are assumed to generate a force dipole on the surrounding fluid. In three dimensions, each particle is assumed to generate a force $-f$ in each direction $\pm \p$, and since the particle has length $2a$, we will model the dipole strength as $\sigma=2a f$. The total stress generated by the active particles in a given volume of fluid is written as $\bsigma^*_a(a\x,Tt)=(K/a^2)\bsigma_a(\x,t)$, where (integrating by parts),
\begin{gather}
\nabla'\cdot \bsigma^*_a(a\x,Tt) = \int_{\mathcal{D}^*}\int_{\mathbb{S}^{2}}\sigma \p\p \cdot \nabla'\delta^*(a\b{x}-a\b{x}_0)\psi^*(a\x_0, \p, Tt) \,d\Omega\,(a^3 dV_0)\\
=\nabla'\cdot \int_{\mathbb{S}^{2}}\sigma \p\p \psi^*(a\x, \p, Tt) \,d\Omega,
\end{gather}
so that the dimensionless active stress is given by
\begin{gather}
\bsigma_a(\x,t) = \Er \volfrac \langle \p\p \rangle = \Er \volfrac  c\D.
\end{gather}
Here we have defined the active Ericksen number, which depends on the particle concentration:
\begin{gather}
\Er = \frac{a^2\sigma}{K v}=\frac{2f}{K}\frac{a^3}{v},
\end{gather}
recalling that $v$ is the particle volume and $\volfrac$ is the particle volume fraction. With $\Er<0$, the active particles are `pushers', and with $\Er>0$, the particles are `pullers'.

\section{Comparison of 2D moment equations and reduced 1D system}

\begin{figure}[ht]
    \centering
    \includegraphics[width=.72\textwidth]{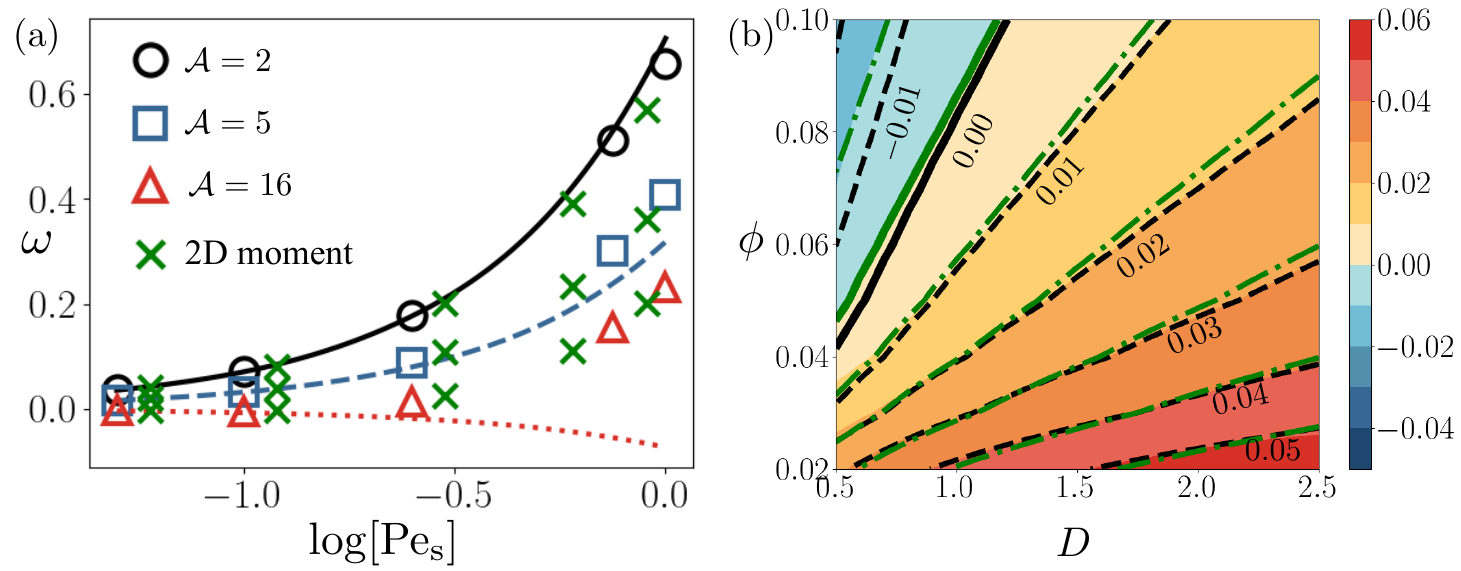}
    \caption{(a) Traveling wave speeds, computed using the 1D equations (Eqs.~16 \& 19 in the main text) are shown as symbols for three different activity parameters, $\Act$, and those computed using the 2D moment equations (Eqs. 14, 15, and 8 in the main text) are shown as green crosses. The curves are the analytical estimates included in the main text. (b) A comparison of the contour plot in Fig.~5c in the main text with curves generated by simulations of the 2D moment equations, denoted with green dot-dash lines. }
    \label{fig:FigureS1}
\end{figure}

Figure~\ref{fig:FigureS1}a shows a comparison of the wave speeds in the traveling arrested states using different methods of computing them, along with the theoretical prediction provided in the main text. Wave speeds computed using the 1D equations (16 \& 19 in the main text) are shown along with the predicted curves from the theoretical prediction, just as they are in Fig.~5a in the main text. But here as well, we include the computed values using the 2D moment equations (Eqs.~14-15, along with Eq.~8 in the main text) as green crosses. Figure~\ref{fig:FigureS1}b performs the same comparison, but for Fig.~5c in the main text. 

\newpage




\section{Movie descriptions}
\begin{itemize}
\item Movie 1: Relaxation of random initial data to equilibrium for sufficiently small active Ericksen number (here, $(\Er, \Act)=(-0.1, 0.06)$) [kinetic theory].

\item Movie 2: The first flowing arrested state for immotile particles, emerging beyond a critical active Ericksen number (or particle concentration); $(\Er, \Act) =(-8, 5)$ (Fig.~1b in the main text) [kinetic theory]. 

\item Movie 3: A fully two-dimensional flowing arrested state, for immotile particles with $(\Er, \Act) = (-16 ,10)$ (Fig.~1c in the main text) [kinetic theory]. 

\item Movie 4: A traveling wave in a system of motile particles ($\V =0.06$) with $(\Er, \Act) = (-3 ,2)$ (Fig.~1d in the main text) [moment equations].

\item Movie 5: A retrograde traveling concentration wave with motile particles ($\V =0.06$) at larger active Ericksen number, with $(\Er, \Act) = (-7 ,37)$ (Fig.~1e in the main text) [moment equations].

\item Movie 6: A traveling periodic `thrashing' mode emerges at larger swimming speeds ($\V =1$), with  $(\Er, \Act) = (-15 ,10)$ [moment equations].

\item Movie 7: At large particle activity, $\Er=-70$, and small anchoring strength, $\W=0.1$, a chaotic flowing state emerges as in Newtonian fluids. At large anchoring strength, $\W=20$, the system settles into an arrested (flowing) state, with high wavenumber distortions and pseudo-defects as in active nematics [kinetic theory].

\item Movie 8: Exploring the role of the anchoring strength, $\W$. A transition from the first arrested state to the second is observed for $\W \in \{0.02, 0.05, 0.1, 0.1, 1.0\}$ with fixed $(\Er, \eta, \volfrac, D, d) = (-8,1,0.02,0.01,0.01)$ (see Fig.~4 in the main text) [kinetic theory],

\end{itemize}

\end{document}